# A novel quantitative indicator of the left ventricular contraction based on volume changes of the left ventricular myocardial segments

Mersedeh Karvandi, MD, Saeed Ranjbar, Ph.D.

Research institute: Taleghani Hospital, Shahid Beheshti University of Medical Sciences,Tehran, Iran

## Abstract:

**Background:**

Ejection fraction (EF) is commonly measured by echocardiography, by dividing the volume ejected by the heart (stroke volume) by the volume of the filled heart (end-diastolic volume). Utilizing volume changes of left myocardial segments per a cardiac cycle, physical laws and mathematical equations specific echocardiographic data, this paper serves to generalize EF by a novel parameter over the time that it can make available, more detailed, valuable and practical information to fully describe the left ventricular (LV) contractility function.

**Methods and Results:**

Patients who underwent clinically-directed standard transthoracic echocardiography using 2D conventional echocardiography machines armed to measuring strain components, were asked to estimate displacements and longitudinal, radial and circumferential strains for each LV echocardiographic segments per a cardiac cycle. Volume fractional changes of the LV echocardiographic segments are expanded based on their strain components over the time. Ejected blood volume fraction induced by a left myocardial sample, is computed within a cardiac cycle. Total fraction of the ejected blood volume in the left ventricular cavity was obtained by integrating over the times and LV myocardial segments. EF is an especial value of this total fraction at the end systolic time.

**Conclusion:**

The common measurement of EF is only based on LV cavity volumes at the end diastolic and systolic phases. These findings lead to determine detailed aspects of the left ventricular contraction. This generalized parameter has important implications to give the real value of EF in the sever Mitral valve regurgitations.

**Introduction:**

The systolic phase of the left ventricle is a complex step that comprise a coordinate contraction of subendocardial, midwall, and subepicardial muscle fibers. These fibers are arranged in a complex, helical manner. Midwall fibers are oriented circumferentially; contraction of these fibers mostly contributes to a decrease in the minor axis of the left ventricle and is responsible for generation of much of the ejection fraction (EF). Longitudinally oriented fibers in subendocardium and subepicardium contribute to shortening of the long axis of the ventricle, also contributing to EF. Resulting in the circumferential and longitudinal strains in normal cases, myocardial fibers movements are started from the posterior-basal region of the heart, continues through the LV free wall, touches the septum, rings around the apex, rises, and ends at the superior-anterior edge of LV. [1,2]

Contractile function is a general term to describe changes in performance or function that could be affected by changes in inotropic state, fiber length, or load. Thus, a change in the EF might best be classified as a change in contractile function; this term is often used interchangeably with systolic function. To fully describe the left ventricle it is necessary to define several additional terms that refer to its mechanical properties.

In scenario with these shortening and twisting deformation, wall thickening contributes volume displacement and creation of the EF. For a LV myocardial segment (Figure 1), the volume changes of it within a cardiac cycle, might be a quantitative marker of regional function of the LV myocardium. Utilizing physical laws and mathematical equations, specific echocardiographic data can make available more detailed, valuable and practical information for the volume change measurements.
This study serves to provide a real time parameter of regional LV contractility even more than one for global function of the left ventricular mechanisms.

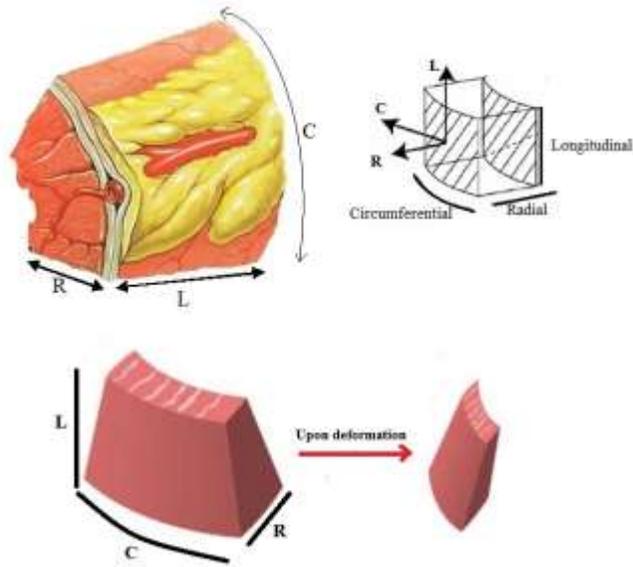

**Figure 1:** An anatomical LV myocardial sample (from the top at the left side), A local coordinate attached to it (from the top at the right side), A LV myocardial segment upon deformation (the bottom). L: Longitudinal direction; R: Radial direction; C: Circumferential direction.

## Methods and Results:

Patients who underwent clinically-directed standard transthoracic echocardiography using 2D conventional echocardiography machines armed to measuring strain components (Longitudinal, Radial and Circumferential strains), were asked to acquire datasets from apical 4-chamber (4-C), 2-chamber (2-C) and short axis views. An expert investigator was requested to first qualitatively estimate displacements and longitudinal, radial and circumferential strains for each LV echocardiographic segments per a cardiac cycle (Figure 2).

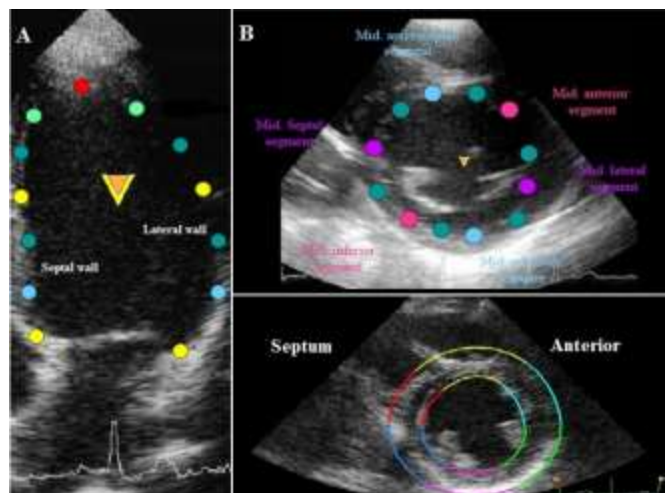

**Figure 2:** Illustrations showing a myocardial segment model of the left ventricular wall in different views.

Volume fractional changes of the LV echocardiographic segments are expanded based on their strains indices over the time. For a muscle myocardial sample of the left ventricle, we named its volume by V(t) at time t.

$\frac{\nabla V(t)}{V(t_d)} = \frac{[V(t)-V(t_d)]}{V(t_d)}$ which $t_d$ is the initial time at the end diastole. $V(t) = L(t)R(t)C(t)$ and $V(t_d) = L(t_d)R(t_d)C(t_d)$. Let $\varepsilon_L(t) = \frac{[L(t)-L(t_d)]}{L(t_d)}$; $\varepsilon_R(t) = \frac{[R(t)-R(t_d)]}{R(t_d)}$; $\varepsilon_C(t) = \frac{[C(t)-C(t_d)]}{C(t_d)}$ are longitudinal strain, radial strain and circumferential strain respectively at time t. We can easily rewrite the $\frac{\nabla V(t)}{V(t_d)}$ in terms of strain components by the following way:

$L(t) = L(t_d)\varepsilon_L(t) + L(t_d); R(t) = R(t_d)\varepsilon_R(t) + R(t_d); C(t) = C(t_d)\varepsilon_C(t) + C(t_d)$

$\frac{\nabla V(t)}{V(t_d)} = \frac{[V(t)-V(t_d)]}{V(t_d)} = \frac{[L(t)R(t)C(t) - L(t_d)R(t_d)C(t_d)]}{L(t_d)R(t_d)C(t_d)}$

$\frac{\nabla V(t)}{V(t_0)} = [(L(t_d)\varepsilon_L(t) + L(t_d)) * (R(t_d)\varepsilon_R(t) + R(t_d)) * (C(t_d)\varepsilon_C(t) + C(t_d)) - L(t_d)R(t_d)C(t_d)]/ L(t_d)R(t_d)C(t_d)$

$= \varepsilon_L(t)\varepsilon_R(t)\varepsilon_C(t) + \varepsilon_L(t)\varepsilon_R(t) + \varepsilon_L(t)\varepsilon_C(t) + \varepsilon_R(t)\varepsilon_C(t) + \varepsilon_L(t) + \varepsilon_R(t) + \varepsilon_C(t)$

$\frac{\nabla V(t)}{V(t_0)} = \varepsilon_L(t)\varepsilon_R(t)\varepsilon_C(t) + \varepsilon_L(t)\varepsilon_R(t) + \varepsilon_L(t)\varepsilon_C(t) + \varepsilon_R(t)\varepsilon_C(t) + \varepsilon_L(t) + \varepsilon_R(t) + \varepsilon_C(t)$

The above formula states the volume fractional changes as a function of longitudinal, radial and circumference strains. Let **DV(t)** is the displacement of the volume $V(t)$ **at time t**. using hemodynamics law in physic (Figure 3), the ejected blood volume fraction at time t induced by $\nabla V(t)$ is $\frac{\nabla V(t)}{V(t_d)} * DV(t)$ at the time t.

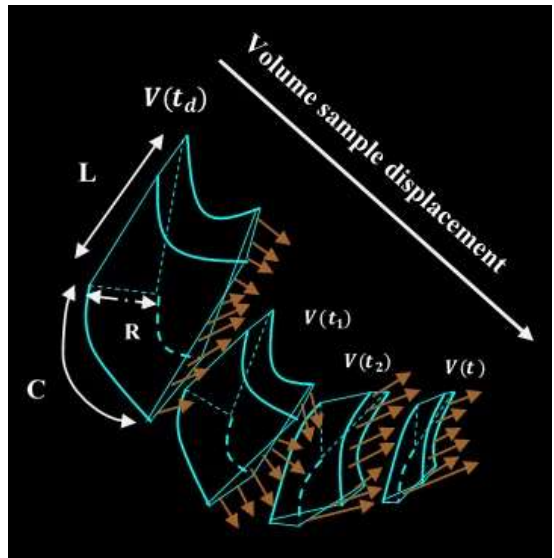

**Figure 3:** Brown vectors represent the impact of the myocardial volume changes to shift the blood fluid in the LV cavity within a cardiac sample.

Finally, the total ejected blood volume fraction (TEBVF) by the myocardial segments in the left ventricular cavity, is obtained by the integration from $\frac{\nabla V(t)}{V(t_d)} * DV(t)$ over the times and LV myocardial segments.

$$\mathbf{TEBVF(t)} = \int_{t_d}^{t} \int_{V \text{ runs through the LV myocaldial samples}} \frac{\nabla V(t)}{V(t_d)} \times DV(t)$$

$$= \int_{t_d}^{t} \int_{V \text{ runs through the LV myocaldial samples}} [\varepsilon_L(t)\varepsilon_R(t)\varepsilon_C(t) + \varepsilon_L(t)\varepsilon_R(t) + \varepsilon_L(t)\varepsilon_C(t) + \varepsilon_R(t)\varepsilon_C(t) + \varepsilon_L(t) + \varepsilon_R(t) + \varepsilon_C(t)] \times DV(t)$$

Such that $t \epsilon [t_d, t_s]$ and $t_s$ is the time at the end systolic phase.

**Discussion:**

1- The new introduced volume strain $VS(t) = \frac{\nabla V(t)}{V(t_d)}$ of a myocardial segment of the left ventricle over the time is a regional computable indicator based on strain components of the LV mechanism. And the global volume strain $\mathbf{GVS(t)} = \sum_V \frac{\nabla V(t)}{V(t_d)}$ (where V runs through all LV's volume samples) is a parameter to fully describe the global LV mechanic at time t. $\mathbf{TVS(t)} = \int_{t_d}^{t} \sum_V \frac{\nabla V(t)}{V(t_d)}$ is the total volume strain of the left ventricular myocardium over the time.

2- We obtained several left ventricular contractility variables $\mathbf{LVC(t)} = \frac{\nabla V(t)}{V(t_d)} \times DV(t)$, $\mathbf{GLVC(t)} = \sum_V \frac{\nabla V(t)}{V(t_d)} \times DV(t)$ and $T\mathbf{EBVF(t)} = \int_{t_d}^{t} \sum_V \frac{\nabla V(t)}{V(t_d)} \times DV(t) = \int_{t_d}^{t} \sum_V \varepsilon_L(t)\varepsilon_R(t)\varepsilon_C(t) + \varepsilon_L(t)\varepsilon_R(t) + \varepsilon_L(t)\varepsilon_C(t) + \varepsilon_R(t)\varepsilon_C(t) + \varepsilon_L(t) + \varepsilon_R(t) + \varepsilon_C(t) \times DV(t)$ in the LV cavity based on the volume strains, the global volume strains and the total volume strains of the left ventricular myocardial samples.

3 - TEBVF is realized as a function such that has the EF as an especial value ( $EF = TEBVF(t_s)$ ).

4 - The TEBVF is an independent parameter of the heart load and heart tethering and it can be estimated automatically by the formula.

5- It may introduce a non-invasive method to test the viability in patients with ischemic heart diseases.

6- As an application, TEBVF($t_s$) gives the real value of EF and this value has a main role to make an exact decision before surgical tasks. For a patient with sever mitral valve regurgitation and EF 40%, a MVR or a mitral valve repair is reported for a cardiac surgeon. Since a fraction of the blood volume was regurgitated to the left atrium, TEBVF($t_s$) with the value 10% candidates the patient for a heart transplantation. For another example, a patient with the aortic valve stenosis and high blood pressure toward the left ventricular hypertrophy it might be possible to use TEBVF($t_s$) for the real value of EF.

7- According to the continuous geometry, [3] a 3D reconstructive image of the LV from conventional 2D echocardiographic images makes a border tracking of the LV. First, all 2D echocardiographic images in all different views are acquired from a 2D echocardiography machine; then region of interests as the considered segments would be determined by the speckle tracking software. [4] A range of the mesh screens can be achieved alongside these images by using the 4D LV-analysis function software (TomTec Imaging Systems GmbH, Munich,Germany) in which each network is included a LV segment. These networks (mesh) are connected together and a 3D mathematical model of the left ventricle would be created. Since this 3D LV reconstruction is based on manipulation of 2D images, our formulas can be easily applied to this 3D LV modeling to gain as one of the exact approaches for estimations of introduced variables.

8- Christian Knackstedt et al. [5] introduced a fully automated versus standard tracking of left ventricular (LV) ejection fraction (EF) and global longitudinal strain (GLS), using the machine learning-enabled image analysis (AutoLV, TomTec-Arena 1.2, TomTec Imaging Systems, Unterschleissheim, Germany). Voigt JU et al. [4] and Farsalinos KE et al. [6] and Thomas H. Marwick et al. [7] have recently also studied the newer algorithms that rely primarily on speckle tracking and artificial intelligence in tracking techniques to detect and contour the endocardium and cardiac cycle and they run the LS, EF and analogous measurements. One questionable concern that rises at their results as to whether these progresses in the measurement of strain will allow GLS to displace EF. Although, GLS is the most strong of the deformation markers, reproducing the promising effect of averaging when individual measurements are subject to noise, we need a computable indicator of regional function even more than one for global function. The new provided quantitative marker (TEBVF(t)) at this article suggests the same needed calculable indicator to analysis the left ventricular function.

**Conclusion:**

The TEBVF parameter has important implications to give the real value of EF in the sever Mitral valve regurgitations and makes the lack of advance and expensive echocardiographic software and will allow TEBVF to replace GLS and EF.

**Conflict of interest:**

There is no conflict of interest.

**References:**

1- Kocica MJ, Corno AF, Carreras-Costa F, Ballester-Rodes M, Moghbel MC, Cueva CNC, Lackovic V, Kanjuh VI, Torrent-Guasp F. The helical ventricular myocardial band: global, three-dimensional, functional architecture of the ventricular myocardium. Eur J Cardiothorac Surg 2006;29:S21-40.

2- 4D LV-Analysis- TomTec - Imaging Systems. http://www.tomtec.de/end_users/4d_echo/4d_lv_analysisc.html (March 2015, date last accessed).

3- Neumann JV, Continuous Geometry, Princeton Univ. Press, Princeton 1960; 22: 92-100.

4- Voigt JU, Lysyansky P, Marwick TH, Houle H, Baumann R et al. Definitions for a common standard for 2D speckle tracking echocardiography: consensus document of the EACVI/ASE/Industry Task Force. J Am Soc Echocardiogr 2015; 28:183-193.

5- Knackstedt C, Bekkers S. C.A.M., Schummers G. et al. Fully Automated Versus Standard Tracking of Left Ventricular Ejection Fraction and Longitudinal Strain. J Am Coll Cardiol 2 0 1 5; 66 : 1 4 5 6 – 6 6

6- Farsalinos KE, Daraban AM, Ulnu S,Thomas JD, Badano LP, Voigt JU. Head-to-Head comparison of global longitudinal strain measurements among nine different vendors the EACVI/ASE inter-vendor comparison study. J Am Soc Echocardiogr 2015; 28:1171-1181.

7-Marwick TH. The Clinical Application of Strain: Raising the Standa. J Am Soc Echocardiogr 2015 ;28:1182-1183